\documentclass[aps,prl,twocolumn,10pt,showpacs,superscriptaddress,groupedaddress,longbibliography]{revtex4-1} 
\pdfoutput=1

\usepackage{graphicx}        
\usepackage{dcolumn}        
\usepackage{bm}             
\usepackage{amssymb}
\usepackage{amsmath}
\usepackage{epstopdf} 
\usepackage{color}
\definecolor{korr_26Apr}{rgb}{0,0,0} 
\definecolor{red}{rgb}{1,0,0}

\begin{document}

\widetext

\title{Local Rheology Relation with Variable Yield Stress Ratio across Dry, Wet, Dense, and Dilute Granular Flows}
\author{Thomas P\"ahtz$^{1,2}$}
\email{0012136@zju.edu.cn}
\author{Orencio Dur\'an$^3$}
\author{David N. {de Klerk}$^{4,5}$}
\author{Indresan Govender$^6$}
\author{Martin Trulsson$^7$}
\affiliation{1.~Institute of Port, Coastal and Offshore Engineering, Ocean College, Zhejiang University, 310058 Hangzhou, China \\
2.~State Key Laboratory of Satellite Ocean Environment Dynamics, Second Institute of Oceanography, 310012 Hangzhou, China \\
3.~Department of Ocean Engineering, Texas A\&M University, College Station, Texas 77843-3136, USA \\
4.~Centre for Minerals Research, University of Cape Town, Private Bag Rondebosch, 7701, South Africa \\
5.~Department of Physics, University of Cape Town, Private Bag Rondebosch, 7701, South Africa \\
6.~School of Engineering, University of KwaZulu-Natal, Glenwood, 4041, South Africa \\
7.~Theoretical Chemistry, Department of Chemistry, Lund University, P.O. Box 124, SE-221 00 Lund, Sweden}

\begin{abstract}
Dry, wet, dense, and dilute granular flows have been previously considered fundamentally different and thus described by distinct, and in many cases incompatible, rheologies. We carry out extensive simulations of granular flows, including wet and dry conditions, various geometries and driving mechanisms (boundary driven, fluid driven, and gravity driven), many of which are not captured by standard rheology models. For all simulated conditions, except for fluid-driven and gravity-driven flows close to the flow threshold, we find that the Mohr-Coulomb friction coefficient $\mu$ scales with the square root of the local P\'eclet number $\mathrm{Pe}$ provided that the particle diameter exceeds the particle mean free path. With decreasing $\mathrm{Pe}$ and granular temperature gradient $M$, this general scaling breaks down, leading to a yield condition with a variable yield stress ratio characterized by $M$.
\end{abstract}

\maketitle
Reliable large-scale simulations and thus predictions of geophysical and industrial processes require a deep understanding of the continuum properties of granular flows. However, existing theories of the granular flow rheology are limited to small subsets of the physical conditions under which such processes can occur. For example, although geophysical granular flows are often wet (i.e., significantly affected or driven by ambient fluid)~\cite{CourrechduPontetal03,HoussaisJerolmack17,Delannayetal17} and consist of coexisting dense (liquidlike) and dilute (gaslike) flow layers~\cite{Borzsonyietal09,HolyoakeMcElwaine12,Broduetal13,Broduetal15,Delannayetal17}, even understanding comparably simple dry, dense-only or dilute-only flows has remained a major challenge~\cite{MiDi04,Andreottietal13,Jop15,Kumaran15}. 

Existing rheologies for noncohesive and nonquasistatic flows of sufficiently hard granular particles can be classified in terms of the particle volume fraction $\phi$ (the fraction of space covered by particles), particle-fluid-density ratio $s\equiv\rho_p/\rho_f$, and Stokes number $\mathrm{St}\equiv\rho_p\dot\gamma d^2/\eta_f$, where $d$ is the particle diameter, $\dot\gamma$ the granular shear rate, and $\eta_f$ the fluid viscosity. Dilute, dry flows ($\phi\lesssim0.5$, $s\rightarrow\infty$, $\mathrm{St}\rightarrow\infty$) have been described by the kinetic theory of dry granular gases~\cite{SelaGoldhirsch98,SahaAlam14,SahaAlam16,GarzoDufty99}, dense, dry flows ($\phi\gtrsim0.5$, $s\rightarrow\infty$, $\mathrm{St}\rightarrow\infty$) by the local viscoplastic rheology~\cite{Jopetal06} and its nonlocal extensions~\cite{KamrinKoval12,Bouzidetal13,Bouzidetal15,ZhangKamrin17}, dense solid-liquid suspensions ($\phi\gtrsim0.5$, $s\simeq1$, variable $\mathrm{St}$) by different (partially incompatible~\cite{GuazzelliPouliquen18}) viscoinertial rheologies~\cite{Boyeretal11,Trulssonetal12,NessSun15,NessSun16,Amarsidetal17,DeGiulietal15}, and sediment transport driven by liquids (variable $\phi$, $1<s<3$, variable $\mathrm{St}$) by modified viscoplastic or viscoinertial rheologies~\cite{Houssaisetal15,Houssaisetal16,Maurinetal16}. Furthermore, different rather complex and controversial approaches exist to extend kinetic theory to solid-gas suspensions~\cite{Garzoetal12,Chamorroetal15,SahaAlam17,Alametal19} or the dense regime~\cite{ChialvoSundaresan13,Vescovietal14,BerziVescovi15}.

Here we show that, despite their fundamental differences, granular flows from the entire phase space $(\phi,s,\mathrm{St})$ actually obey a common scaling law for the Mohr-Coulomb friction coefficient $\mu$, the knowledge of which is essential for any rheological description.

\begin{figure}[htb]
 \begin{center}
  \includegraphics[width=1.0\columnwidth]{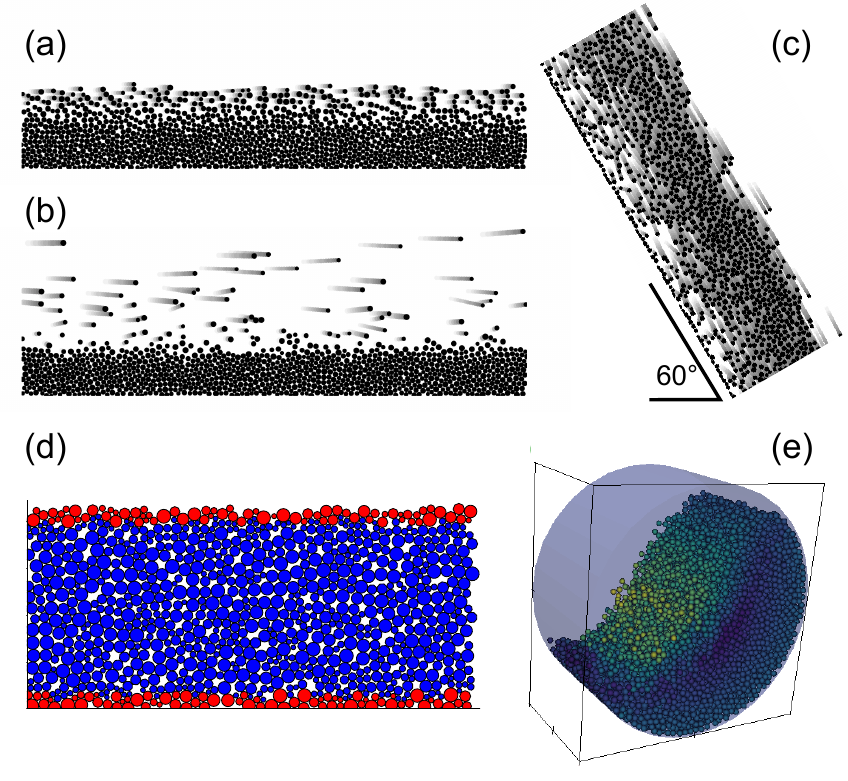}
 \end{center}
 \caption{Visualizations of numerical simulations of (a) and (b) sediment transport driven by various Newtonian fluids, (c) rapid gravity-driven flows in ambient static air, (d) uniformly sheared flows, and (e) rotating drum flows.}
\label{Screenshots}
\end{figure}

We carry out discrete element method-based simulations of granular flows for a variety of geometries and driving mechanisms (Table~\ref{SimulationSummary} and Fig.~\ref{Screenshots}), which cover the entire phase space: (i) two-dimensional sediment transport driven by a large variety of Newtonian fluids; (ii) two-dimensional rapid gravity-driven flows in ambient static air of varying viscosity, many of which are highly convective (e.g., they can exhibit a strong kinetic heat transfer normal to the flow direction) and/or ``supported''~\cite{Broduetal15}; (iii) two-dimensional uniformly sheared viscous suspensions in density-matched fluid of varying viscosity; (iv) two-dimensional dry uniform shear flows; (v) three-dimensional rotating drum flows lubricated by a density-matched fluid; and (vi) a three-dimensional dry rotating drum flow. Among these flows, rapid gravity flows and rotating drum flows are known to elude the description by standard rheology models~\cite{Cortetetal09,Borzsonyietal09,HolyoakeMcElwaine12,Broduetal13,Broduetal15,Govender16}. In all simulations, contacting particles interact via normal repulsion (restitution coefficient $e$, modeled through viscous damping), governed either by a linear or Hertzian law, and tangential friction (contact friction coefficient $\mu_c$, Table~\ref{SimulationSummary}). Details are described below.
\begin{table}[htb]
    \begin{tabular}{ l | c | c }
    \hline
    \hline
     Flow geometry & Driven by & Contact model ($e$, $\mu_c$) \\ 
    \hline
		\hline
		 Sediment transport (2D) & Fluid & Linear (0.9, 0.5) \\
		 Rapid gravity flows (2D) & Gravity & Linear (0.9, 0.5) \\
		 Sheared suspensions (2D) & Boundary & Linear (0.1, 0.4) \\
		 Dry shear flows (2D) & Boundary & Linear (0.1, 0.4) \\
		 Lubricated drum flows (3D) & Boundary & Hertz (0.5, 0.5) \\
		 Dry drum flow (3D) & Boundary & Hertz (0.5, 0.5) \\
		 \hline
    \end{tabular}
		\caption{Summary of simulated granular flows.}
		\label{SimulationSummary}
\end{table}

\textit{Sediment transport and gravity flows.---}The numerical model couples a discrete element method for the particle motion (stiffness $k=5000mg/d$) under gravity, buoyancy, and fluid drag with a continuum Reynolds-averaged description of hydrodynamics (described in detail and/or validated in Refs.~\cite{Duranetal11,Duranetal12,Duranetal14b,PahtzDuran17,PahtzDuran18a,PahtzDuran18b}). Spherical particles ($\sim 10^4$) with mild polydispersity are confined in a quasi-two-dimensional, vertically infinite domain of length $\sim 10^3 d$ with periodic boundary conditions in the flow direction. For gravity flows, the ambient fluid is kept static. 

Simulations are carried out for varying density ratio $s$, Galileo number $\mathrm{Ga}\equiv\rho_f\sqrt{(s-1)gd^3}/\eta_f$, Shields number $\Theta\equiv\tau_f/[(\rho_p-\rho_f)gd]$, and inclination angle $\alpha$, where $g$ is the gravitational constant and $\tau_f$ is the bed fluid shear stress. For gravity flows, we simulate conditions with $s=2000$, $\mathrm{Ga}\in[2,100]$, $\Theta=0$, and $\alpha$ between the flow threshold and $60^\circ$. For sediment transport, we simulate conditions with $s\in[2.65,2000]$, $\mathrm{Ga}\in[0.1,100]$, $\alpha=0$, and $\Theta$ above the flow threshold, which correspond to five different transport regimes (Table~\ref{SedimentTransportConditions})~\cite{PahtzDuran18a}. Following the symmetry along the flow direction, simulation data are averaged over horizontal layers of variable thickness depending on the particle volume fraction~\cite{Duranetal12}.
\begin{table}[htb]
    \begin{tabular}{ p{0.49\columnwidth} | p{0.49\columnwidth} }
     \hline
     \hline
      Sediment transport regime & Condition \\ 
     \hline
	   \hline
    	 Viscous bedload transport & $\sqrt{s}\mathrm{Ga}<20$ \\
		 \hline
		 Turbulent bedload transport & $\sqrt{s}\mathrm{Ga}\geq20\land s<10$ \\
		 \hline
		 Bedload-saltation transition & $20\leq\sqrt{s}\mathrm{Ga}<80\land s\geq10$ \\
		 \hline
		 Viscous saltation transport & $\sqrt{s}\mathrm{Ga}\geq80\land s\geq10\land \sqrt[4]{s}\mathrm{Ga}<32$ \\
		 \hline
		 Turbulent saltation transport & $\sqrt{s}\mathrm{Ga}\geq80\land s\geq10\land \sqrt[4]{s}\mathrm{Ga}\geq32$ \\
		 \hline
    \end{tabular}
		\caption{Sediment transport regimes~\cite{PahtzDuran18a}.}
		\label{SedimentTransportConditions}
\end{table}

\textit{Uniformly sheared particle and suspension flows.---}The numerical model couples a discrete element method for the particle motion ($k=2000P_{zz}d$) under viscous fluid drag and torque with the Stokes equations for laminar flow (described in detail in Refs.~\cite{Trulssonetal12,Trulssonetal17}). Two-dimensional disks ($\sim 10^3$) with moderate polydispersity are confined within a shear cell composed by two rough walls, created by gluing together two dense layers of grains, with periodic boundary conditions along the flow direction parallel to the walls. The position of the walls is controlled to ensure constant confining pressure $P_{zz}$ and mean shear rate.

Simulations are carried out for varying volume fraction (in the range $\phi>0.24$, where $\phi$ is calculated as $2/3$ of the disk area fraction, like for spheres confined in two dimensions) and two general cases: no ambient fluid (dry condition) and an ambient density-matched liquid with varying dimensionless viscosity ($s=1$, $\eta_f/\sqrt{\rho_pP_{zz}d^2}=[10^{-3},10^{-2},10^{-1},10^0,\infty]$).
\begin{figure*}[htb]
 \begin{center}
  \includegraphics[width=2.0\columnwidth]{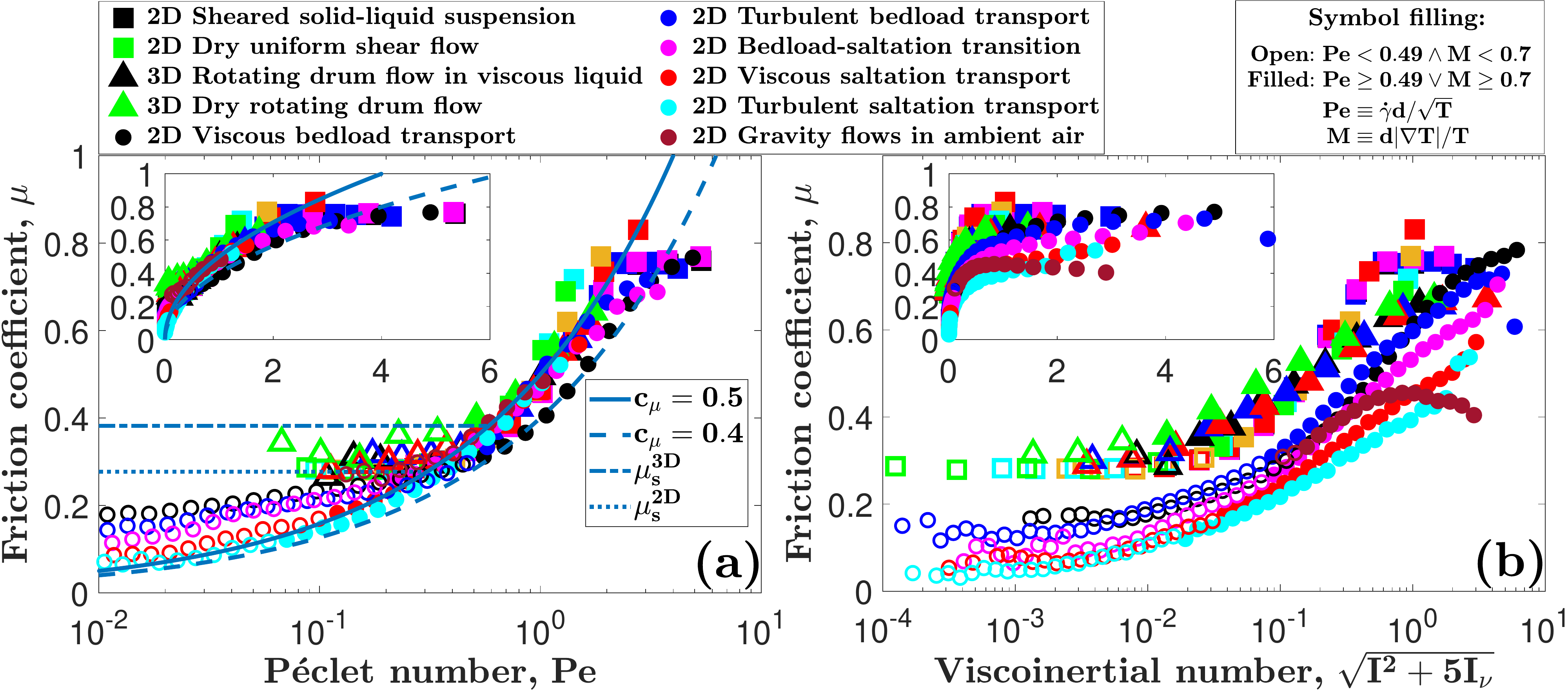}
 \end{center}
 \caption{Mohr-Coulomb friction coefficient $\mu$ vs (a) P\'eclet number $\mathrm{Pe}$ and (b) viscoinertial number $K\equiv\sqrt{I^2+5I_\nu}$ for data from discrete element method-based simulations of various granular flows. The values of $\mathrm{Pe}$, $K$, and $\mu$ depend on the location within the flow. The value of $\mu$ at each location with $\lambda(\phi)<d$ and either $\mathrm{Pe}<0.49$ and $M<0.7$ (open symbols) or $\mathrm{Pe}\geq0.49$ or $M\geq0.7$ (closed symbols) is allocated to the corresponding bin of $\mathrm{Pe}$ or $K$. Each bin consists of data from either a single simulation (rotating drum and uniform flows) or from various simulations of the same regime (sediment transport and gravity flows, see Tables~\ref{SedimentTransportConditions} and S1~\cite{SuppLiquidRheology}). The mean of $\mu$ within each bin is represented by the symbols (for standard deviation, see Fig.~S2 in the Supplementary Material~\cite{SuppLiquidRheology}). For the squares, the color order (green, cyan, orange, red, magenta, blue, black) corresponds to $\eta_f/\sqrt{\rho_pP_{zz}d^2}=[0,10^{-3},10^{-2},10^{-1},10^0,10^1,\infty]$. For the triangles, the color order (green, red, blue, black) corresponds to $\eta_f/(\rho_f\omega d^2)=[0,1/160,1/16,3/16]$.}
\label{Scaling}
\end{figure*}

\textit{Rotating drum flows.---}The numerical model uses a discrete element method for the particle motion ($k=17000mg/d^{3/2}$) under lubrication forces~\cite{NessSun15} and gravity. The contact model employs the LIGGGHTS implementation of Hertzian contacts, which ensures a constant value of $e$~\cite{DiRenzoDiMaio04,DiRenzoDiMaio05}. Spherical monodisperse particles ($\sim 10^4$) are confined within a closed horizontal cylinder (drum) of radius $20d$ and width $20d$ rotating at a constant rate $\omega$.

Simulations are carried out for dry conditions and an ambient density-matched liquid with varying dimensionless viscosity ($s=1$, $\eta_f/(\rho_f\omega d^2)=[1/160,1/16,3/16]$). Simulation data are averaged using an anisotropic Gaussian smoothing function of dimension $3d\times3d\times 20d$.
\begin{figure*}[htb]
 \begin{center}
  \includegraphics[width=2.0\columnwidth]{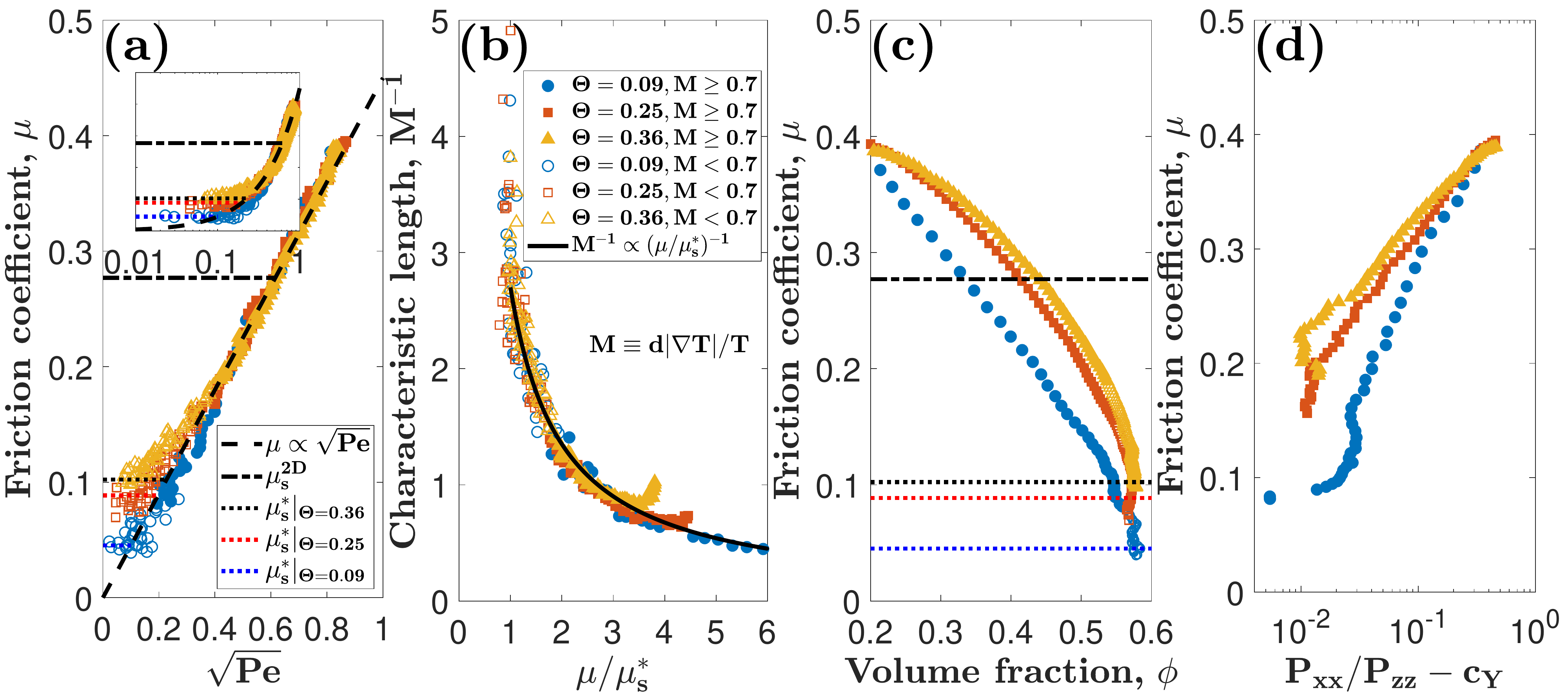}
 \end{center}
 \caption{(a) Mohr-Coulomb friction coefficient $\mu$ vs square root of P\'eclet number ($\sqrt{\mathrm{Pe}}$). Inset: linear-log scale. (b) Characteristic length $M^{-1}$ vs rescaled friction coefficient $\mu/\mu^\ast_s$. (c) $\mu$ vs particle volume fraction $\phi$. (d) $\mu$ vs $P_{xx}/P_{zz}-c_Y$, where $\mathbf{P}=\mathbf{P}^c+\mathbf{P}^k$ and $c_Y=0.98\ne1$ because finite-size effects allow for slight normal stress differences at yielding~\cite{Clarketal18}. Symbols correspond to data with $\lambda(\phi)<d$ from simulations of turbulent saltation transport ($s=2000$, $\mathrm{Ga}=5$) for three different Shields numbers $\Theta$. Filled symbols indicate $\mathrm{Pe}\geq0.49$ or $M\geq0.7$. Open symbols indicate $\mathrm{Pe}<0.49$ and $M<0.7$ [not shown in (d) for visibility reasons].}
\label{Yielding}
\end{figure*}

\textit{General rheology relation.---}Contact ($\mathbf{P}^c$) and kinetic ($\mathbf{P}^k$) granular stresses are calculated from the simulation data using the method given in Ref.~\cite{ArtoniRichard15b}, which ensures that the granular temperature $T=\sum_iP^k_{ii}/(\rho_p\phi D)$ (the root-mean-square of the particle fluctuation velocity), where $D$ is the number of space dimensions, is insensitive to the coarse-graining width. Furthermore, the shear rate $\dot\gamma$ is calculated as the norm of the deviatoric component of the strain rate tensor $\epsilon_{ij}\equiv\partial_i\langle v_j\rangle+\partial_j\langle v_i\rangle$, which reads $\dot\gamma\equiv\lVert\boldsymbol\epsilon^d\rVert=\sqrt{\sum_{ij}\epsilon^d_{ij}\epsilon^d_{ij}/2}$, where $\epsilon^d_{ij}=\epsilon_{ij}-\sum_k\epsilon_{kk}\delta_{ij}/D$. Finally, we calculate the Mohr-Coulomb friction coefficient $\mu\equiv\max_{i,j}|P^c_i-P^c_j|/|P^c_i+P^c_j|$ from the principal components $P^c_i$ of $\mathbf{P}^c$ (a Drucker-Prager definition of $\mu$ and/or a definition that includes kinetic stresses yield slightly but statistically significantly worse results).

For both dilute and dense flow conditions (defined shortly), we find that the friction coefficient $\mu$ is only a function of the P\'eclet number $\mathrm{Pe}\equiv\dot\gamma d/\sqrt{T}$~\cite{ChialvoSundaresan13} and scales as
\begin{equation}
 \mu=c_\mu\sqrt{\mathrm{Pe}}, \label{mu}
\end{equation}
almost everywhere within all simulated conditions [Fig.~\ref{Scaling}(a), closed symbols], except for sediment transport and gravity flows too close to the flow threshold (which are excluded from Fig.~\ref{Scaling}) because of nonlocal effects (see Supplementary Material for details~\cite{SuppLiquidRheology}). The scaling parameter $c_\mu$ slightly depends on the driving conditions, with the smallest value found for viscous bedload transport. For uniformly sheared flows, $c_\mu$ varies significantly with $\mu_c$ ($c_\mu\simeq0.3$ for $\mu_c=0$, $c_\mu\simeq0.55$ for $\mu_c=100$) but not with $e$, even in the extreme cases $\mu_c=100$ (no sliding) and $\mu_c=0$ (always sliding)~\cite{SuppLiquidRheology}. We therefore propose that Eq.~(\ref{mu}) relates the structure anisotropy to $\mu$, where $c_\mu$ partly encodes the anisotropy of the tangential contact force network (which increases with $\mu_c$) and $\mathrm{Pe}$ encodes the anisotropy of the particle assembly (i.e., the unit normal contact vector), relating the diffusion forces toward isotropic configurations to that of the anisotropic compression and extension of the shear~\cite{ChialvoSundaresan13}. This proposition is consistent with dry uniform shear flows, for which both anisotropies correlate with $\mu$~\cite{AzemaRadjai14}. Alternatively, $\mathrm{Pe}$ may also be interpreted as the ratio between the rates of the macroscopic shearing motion ($\dot\gamma$) and microscopic kinetic rearrangements ($\propto\sqrt{T}/d$), which is similar to the original interpretations of the inertial number $I\equiv\dot\gamma d/\sqrt{P^c/\rho_p}$ (where $P^c\equiv\sum_iP^c_{ii}/D$) and viscous number $I_\nu\equiv\eta_f\dot\gamma/P^c$~\cite{Cassaretal05}. The difference is that, for $I$ (and $I_\nu$), the kinetic rearrangement rate has been obtained from assuming a particle fall driven by pressure (and opposed by viscous drag), whereas for $\mathrm{Pe}$, the kinetic rearrangement rate is obtained from the actual relative motion between neighboring particles. The latter rate should be more general, which is supported by the fact that, in contrast to $\mathrm{Pe}$, neither $I$ (even when limited to dry flows), nor $I_\nu$, nor a combination of the two collapse the $\mu$ data [Fig.~\ref{Scaling}(b) and Supplementary Material~\cite{SuppLiquidRheology}], except the viscoinertial number $\sqrt{I^2+5I_\nu}$ for uniformly sheared flows [Fig.~\ref{Scaling}(b)]. Note that a standard nonlocal rheology model~\cite{KamrinKoval12,ZhangKamrin17} and extended kinetic theory also fail to describe our flows~\cite{SuppLiquidRheology}. In particular, in Navier-Stokes order, the latter predicts $\mu\simeq c^{\mathrm{kin}}_\mu\mathrm{Pe}$ for dense flows~\cite{ChialvoSundaresan13,BerziVescovi15} (with a proportionality constant $c^{\mathrm{kin}}_\mu$ that depends on $e$ but not on $\mu_c$), inconsistent with Eq.~(\ref{mu}) and our dry and wet flow data (for dry flows, adding higher-order terms may remedy this discrepancy~\cite{BerziJenkins18}).

We define dilute and dense conditions -- as opposed to rarefied ones -- in terms of the mean free path $\lambda$ through the condition $\lambda<d$, where $\lambda(\phi)=\sqrt{2}d/(12\phi)$ for spherical particles and $\lambda(\phi)=\pi d/(12\sqrt{2}\phi)$ for spheres confined in two dimensions. In fact, we hypothesize that at least some of the few deviations from the scaling in Eq.~(\ref{mu}) at large $\mathrm{Pe}$ [Fig.~\ref{Scaling}(a)] are related to a transition from dilute to rarefied flows at large shear rates, where $\mu$ is limited by the geometrical constraints of high energy collisions~\cite{PahtzDuran18b}.

\textit{Variable yield stress ratio.---}Interestingly, deviations from the scaling in Eq.~(\ref{mu}) at small $\mathrm{Pe}$ (larger-than-predicted values of $\mu$) are well characterized by the dimensionless granular temperature gradient $M\equiv d|\nabla T|/T$ and seem to occur whenever $M\lesssim0.7$ and $\mathrm{Pe}\lesssim0.5$ [Fig.~\ref{Scaling}(a), open symbols]. For uniformly sheared flows (squares in Fig.~\ref{Scaling}), where temperature gradients are negligible ($M\simeq 0$), these deviations are owed to the fact that $\mu$ converges to the yield stress ratio $\mu_s$ ($\mu^{\mathrm{2D}}_s=0.277$~\cite{Trulssonetal12}, $\mu^{\mathrm{3D}}_s=0.382$~\cite{ChialvoSundaresan13}) in the limit of vanishing shear rate. From Eq.~(\ref{mu}), we find that this yield transition in homogeneous flows starts at $\mathrm{Pe}=(\mu_s/c_\mu)^2\approx0.5$.

For inhomogeneous flows, $\mu$ can be substantially smaller than $\mu_s$ when $\mathrm{Pe}\lesssim0.5$ [Figs.~\ref{Scaling}(a) and \ref{Yielding}(a)], at which point deviations from the scaling in Eq.~(\ref{mu}) are controlled by the condition $M\lesssim0.7$. These deviations have several elements in common with a yield transition, as illustrated in Fig.~\ref{Yielding} for turbulent saltation transport, which is a nearly dry granular flow because of a large density ratio ($s=2000$) and large Stokes numbers [$\mathrm{St}\in(10,200)$]. First, $\mu$ seems to converge to a finite value $\mu_s^\ast$ in the limit of vanishing shear rate [Fig.~\ref{Yielding}(a)]. Second, the dimensionless characteristic length $M^{-1}$, associated with spatial changes in the granular temperature, collapses as a function of $\mu/\mu^\ast_s$, peaks with a finite value at $\mu/\mu^\ast_s=1$, and once $\mu/\mu^\ast_s\lesssim1$, the data scatter [Fig.~\ref{Yielding}(b)]. This peak is similar to the divergence of the relaxation length associated with spatial changes of the shear rate ($\dot\gamma$) and granular stresses ($\mathbf{P}^c$) when approaching the yield condition ($\mu\rightarrow\mu_s$) in existing nonlocal rheology models~\cite{Bouzidetal15}. Finally, $\phi$ approaches the packing fraction as $\mu\rightarrow\mu^\ast_s$ [Fig.~\ref{Yielding}(c)]. We thus conclude that $\mu^\ast_s$ is the analog of $\mu_s$ for inhomogeneous flows.

The onset of the yielding transition is thus controlled by the local values of either $\mathrm{Pe}$ for relatively uniform flows (i.e., relatively small $M$) or $M$ for relatively inhomogeneous flows (i.e., relatively large $M$). In the latter case, the yielding transition can expand over a range of $M$ [e.g., $0.37\lesssim M\lesssim0.7$ for turbulent saltation transport, see Fig.~\ref{Yielding}(b)] and coincides with nonlocality in the relation between $\mu$ and $\mathrm{Pe}$. These behaviors are consistent with $\mathrm{Pe}$ playing the role of a granular fluidity. Fluidity inhomogeneities, which are associated with nonlocality~\cite{Bouzidetal15}, then would decrease with increasing $M$ as shear rate inhomogeneities get compensated by temperature gradients, $(\nabla\mathrm{Pe})/\mathrm{Pe}=(\nabla\dot\gamma)/\dot\gamma-(\nabla T)/(2T)$, rendering the rheology local. They are also consistent with vanishing normal stress differences (e.g., $P_{xx}/P_{zz}-1=0$) at $\mu=\mu^{(\ast)}_s$ in the large-system limit~\cite{Clarketal18} [Fig.~\ref{Yielding}(d)] because large velocity and temperature gradients generate normal stress differences~\cite{SelaGoldhirsch98,SahaAlam14,SahaAlam16} and thus prevent the yielding transition. In particular, it seems that, for $M\gtrsim0.7$, the generation of normal stress differences is sufficient to prevent the yielding transition, even for comparably small $\mathrm{Pe}$ and thus $\mu$ [Eq.~(\ref{mu})].

\textit{Conclusions.---}In this study, we have shown that, under certain relatively weak constraints, the Mohr-Coulomb friction coefficient $\mu$ obeys the general scaling $\mu=c_\mu\sqrt{\mathrm{Pe}}$, with the P\'eclet number defined as $\mathrm{Pe}\equiv\dot\gamma d/\sqrt{T}$ (but, in general, disobeys scaling laws from viscoinertial rheology models and extended kinetic theory). This calls for the development of hydrodynamic models for dense granular flows involving granular temperature. Apart from extended kinetic theory, several such models have already been proposed to reproduce several aspects such as hysteresis~\cite{LeeHuang12,DeGiuliWyart17} and drag on an object~\cite{Seguinetal11}. The scaling parameter $c_\mu$ varies with the tangential friction coefficient $\mu_c$ but not with the normal coefficient of restitution $e$, which led us to propose that $\mu=c_\mu\sqrt{\mathrm{Pe}}$ encodes the effects of the anisotropies of the particle assembly ($\mathrm{Pe}$) and tangential contact force network ($c_\mu$) on $\mu$.

The yield stress ratio of granular media, below which granular flows either stop or fundamentally change~\cite{Bouzidetal15}, is often found to be independent of the flow geometry~\cite{Clarketal18}. However, for wind-driven sediment transport, the scaling $\mu\propto\sqrt{\mathrm{Pe}}$ even holds for friction coefficients as low as $\mu\simeq0.08$ and is accompanied by very low yield stress ratios (as low as $\mu^\ast_s\simeq0.04$, reminiscent of vibrated granular flows~\cite{GaudelDeRichter19}), which seem to be caused by relatively large values of the dimensionless temperature gradient $M\equiv d|\nabla T|/T$. Future studies should investigate this link between $\mu^\ast_s$ and $T$ because it may play a role in explaining long-standing open problems, such as the reduction of friction in long-runout landslides~\cite{Legros02,Lucasetal14,Johnsonetal16}.

\begin{acknowledgments}
T. P. acknowledges support from grant National Natural Science Foundation of China (No.~11750410687). M. T. acknowledges funding from the Swedish Research Council (621-2014-4387). Simulations of rotating drums were performed using facilities provided by the University of Cape Town's ICTS High-Performance Computing team (\url{http://hpc.uct.ac.za}). The DEM simulations of uniform flows were performed on resources provided by the Swedish National Infrastructure for Computing (SNIC) at the Centre for Scientific and Technical Computing at Lund University (LUNARC). 
\end{acknowledgments}

%

\section{Supplementary Material}

\setcounter{equation}{0}
\setcounter{table}{0}
\setcounter{figure}{0}
\renewcommand{\theequation}{S\arabic{equation}}
\renewcommand{\thefigure}{S\arabic{figure}}
\renewcommand{\thetable}{S\arabic{table}}

\subsection{Nonlocality in sediment transport and gravity flows}
\begin{figure*}[htb]
 \begin{center}
  \includegraphics[width=2.0\columnwidth]{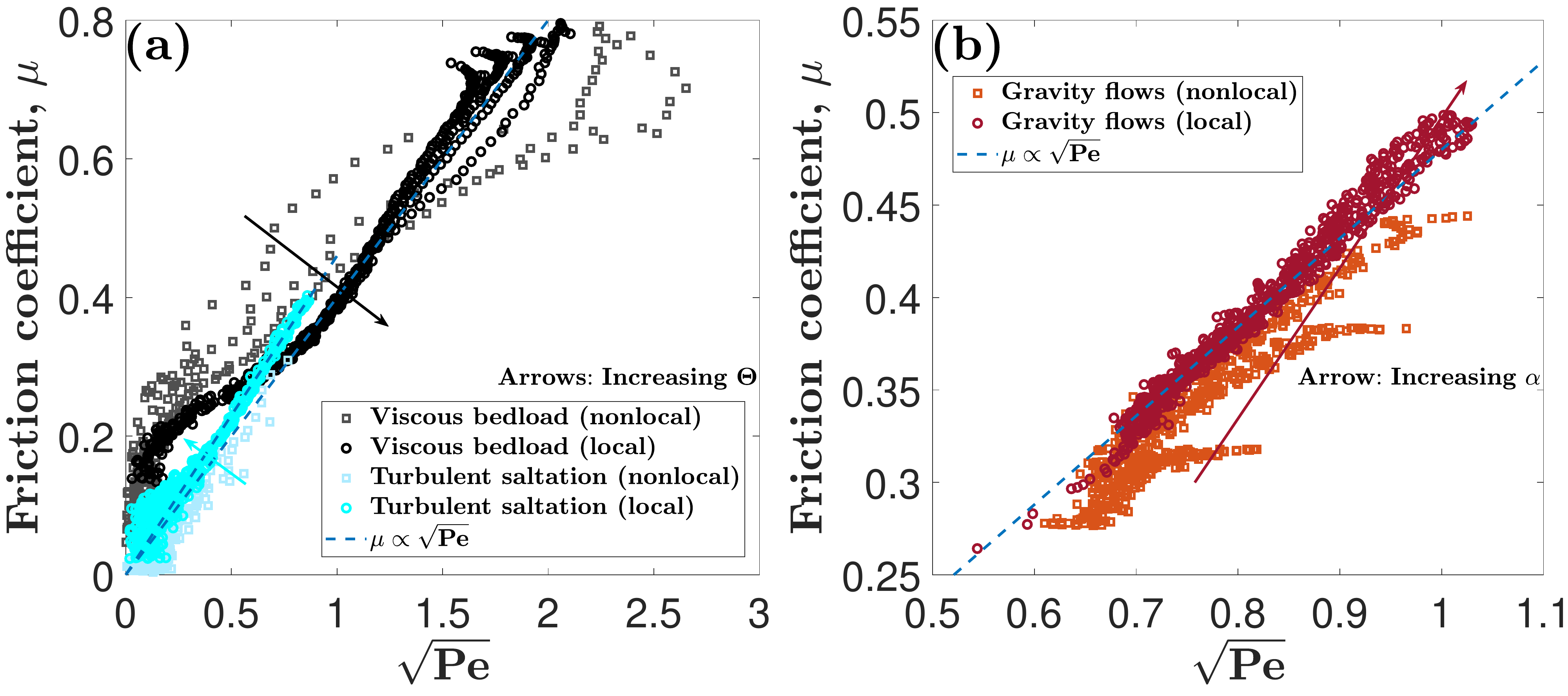}
 \end{center}
 \caption{Mohr-Coulomb friction coefficient $\mu$ vs square root of the P\`eclet number ($\sqrt{\mathrm{Pe}}$) exemplary for (a) viscous bedload transport ($s=2.65$, $\mathrm{Ga}=5$), turbulent saltation transport ($s=2000$, $\mathrm{Ga}=5$), and (b) gravity flows submerged in ambient air ($s=2000$, $\mathrm{Ga}=5$). Symbols correspond to data from our numerical simulations for (a) several Shields numbers $\Theta$ and (b) inclination angles $\alpha$.}
\label{Nonlocality}
\end{figure*}

When the dimensionless fluid shear stress (``Shields number'') $\Theta$ of the fluid driving sediment transport is too close to its value at the flow threshold or when the inclination angle $\alpha$ of the gravity flows in ambient air is too close to its value at the flow threshold, the simulation data do not obey the scaling $\mu\propto\sqrt{\mathrm{Pe}}$ (Fig.~\ref{Nonlocality}), which relates the local friction coefficient $\mu$ to the local P\`eclet number $\mathrm{Pe}$ (i.e., the granular flow rheology is nonlocal). This finding is not surprising because nonlocal effects are already known to be crucial for dry gravity flows (e.g., they are responsible for the stopping angle dependency on the flow thickness \cite{Bouzidetal15}) and because it is already known that sediment transport tends to creep below the surface of the granular bed \cite{Houssaisetal15}, which is also associated with nonlocality \cite{Bouzidetal15}. In particular, for turbulent saltation transport, the granular bed does not flow liquidlike but creeps even relatively far from the flow threshold because the fluid shear stress at the bed surface is insufficient to mobilize particles \cite{Duranetal11}. Only for sufficiently intense conditions (i.e., those in Table~\ref{SimulatedConditions}), when collisions between particles of the rarefied transport layer and granular bed are so frequent that the bed no longer recovers between collisions, does the granular bed flow like a liquid and thus the rheology become local \cite{PahtzDuran18b}.

\subsection{Alternative manner to evaluate the P\'eclet number scaling}
\begin{figure*}[htb]
 \begin{center}
  \includegraphics[width=2.0\columnwidth]{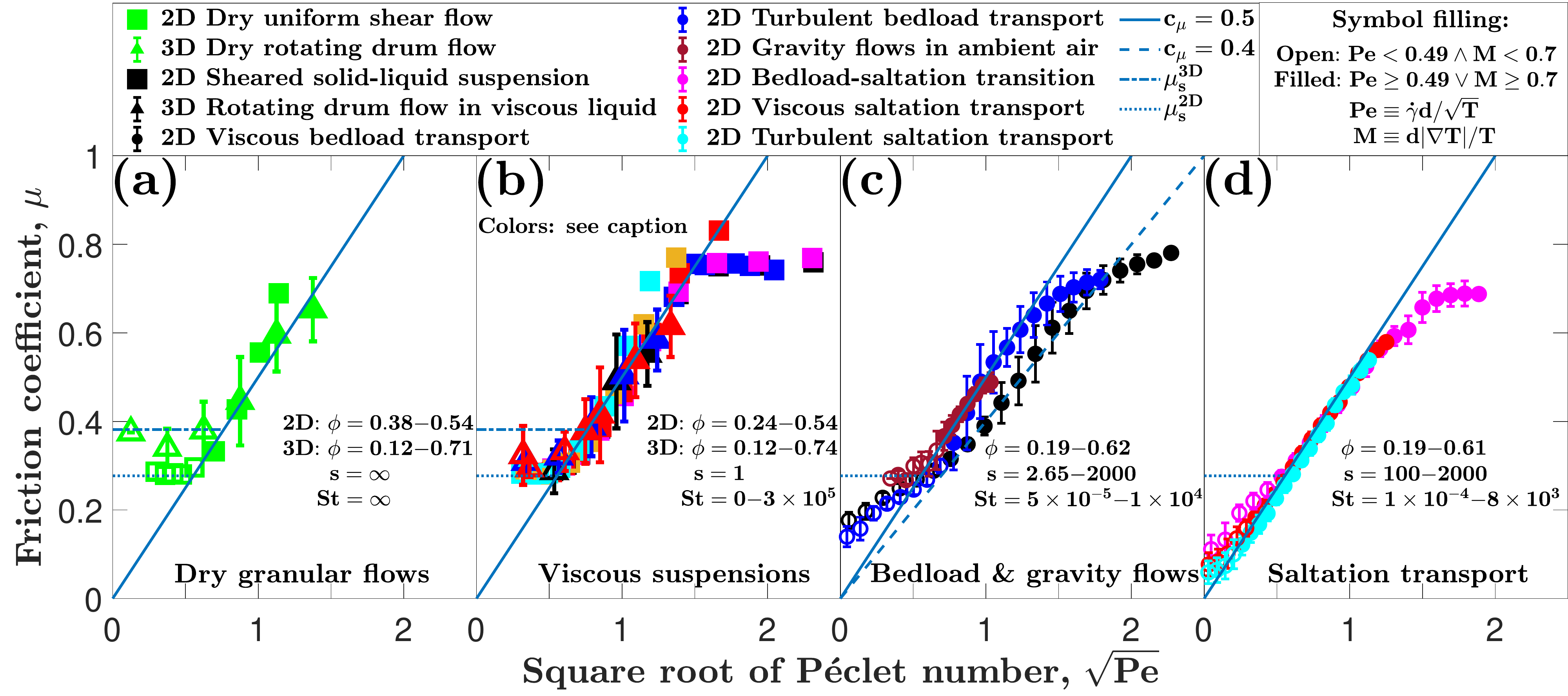}
 \end{center}
 \caption{Mohr-Coulomb friction coefficient $\mu$ vs square root of P\'eclet number ($\sqrt{\mathrm{Pe}}$) for data from discrete element method-based simulations of (a) dry granular flows, (b) viscous suspensions in density-matched liquids, (c) bedload transport and gravity flows, and (d) saltation transport. The values of $\mathrm{Pe}$ and $\mu$ depend on the location within the flow. The value of $\mu$ at each location with $\lambda(\phi)<d$ and either $\sqrt{\mathrm{Pe}}<0.7\land M<0.7$ (open symbols) or $\sqrt{\mathrm{Pe}}\geq0.7\lor M\geq0.7$ (closed symbols) is allocated to the corresponding bin of $\sqrt{\mathrm{Pe}}$. Each bin in (a) and (b) consists of data from a single rotating drum rotating drum or uniform flow simulation, whereas each bin in (c) and (d) consists of data from various simulations of the same regime (Table~\ref{SimulatedConditions}). The mean and standard deviation of $\mu$ within each bin are represented by the symbols and their error bars, respectively. For uniform flows, error bars are usually smaller than symbol size and therefore not shown. For the squares in (b), the color order (cyan, orange, red, magenta, blue, black) corresponds to $\eta_f/\sqrt{\rho_pP_{zz}d^2}=[10^{-3},10^{-2},10^{-1},10^0,10^1,\infty]$. For the triangles in (b), the color order (red, blue, black) corresponds to $\eta_f/(\rho_f\omega d^2)=[1/160,1/16,3/16]$.}
\label{Scaling1}
\end{figure*}

Figure~\ref{Scaling1} shows the scaling $\mu=c_\mu\sqrt{\mathrm{Pe}}$ for the same data as in Fig.~2(a) of the paper, but including the standard deviation for all our simulated flows (for uniform flows, it is usually smaller than the symbol size and therefore not shown).

\subsection{Influence of contact parameters on scaling law}
\begin{figure*}[htb]
 \begin{center}
  \includegraphics[width=2.0\columnwidth]{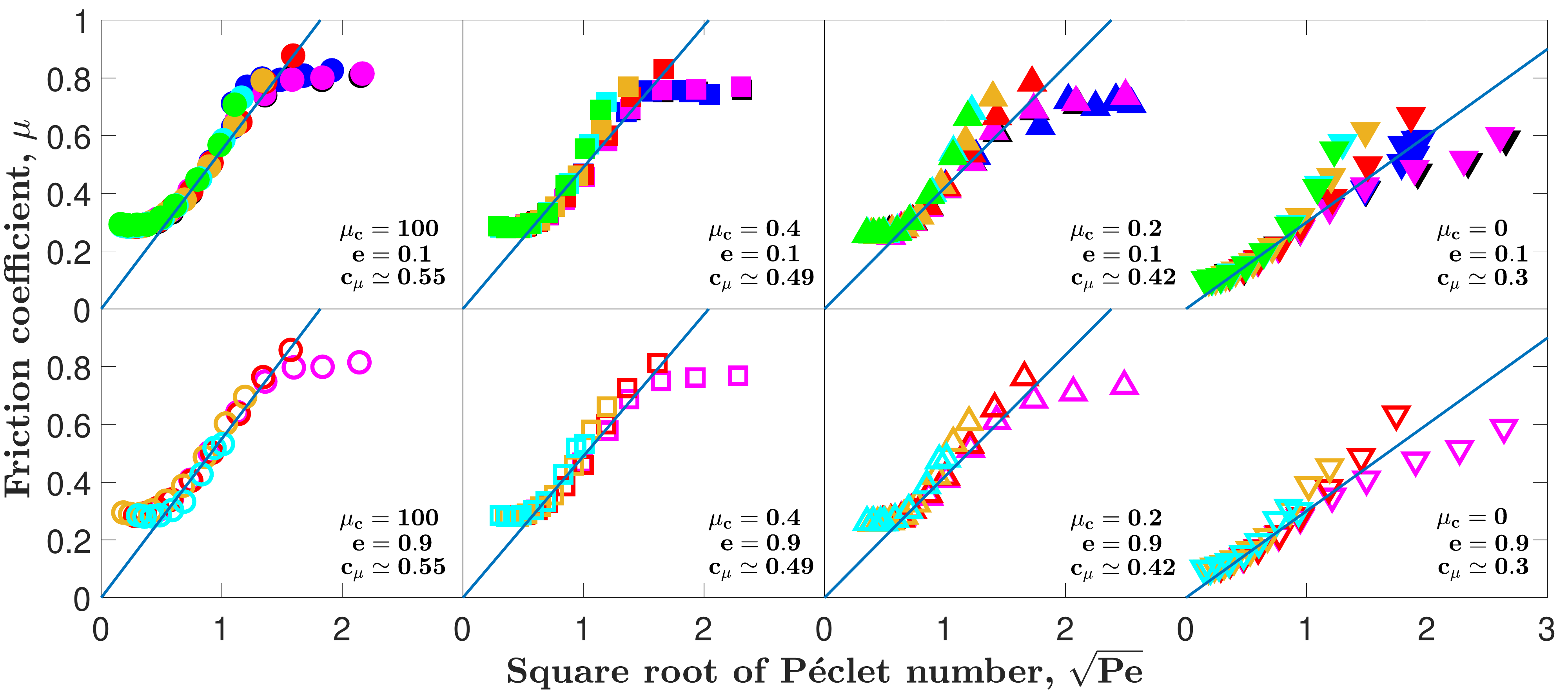}
 \end{center}
 \caption{Mohr-Coulomb friction coefficient $\mu$ vs square root of P\'eclet number ($\sqrt{\mathrm{Pe}}$) for data from discrete element method-based simulations of uniformly sheared particle and suspension flows with varying contact parameters. The color order (green, cyan, orange, red, magenta, blue, black) corresponds to $\eta_f/\sqrt{\rho_fP_{zz}d^2}=[0,10^{-3},10^{-2},10^{-1},10^0,10^1,\infty]$.}
\label{ContactParameters}
\end{figure*}

Figure~\ref{ContactParameters} shows the influence of contact parameters on the scaling $\mu=c_\mu\sqrt{\mathrm{Pe}}$ for uniformly sheared particle and suspension flows. It can be seen that the normal restitution coefficient $e$ does not affect the scaling parameter $c_\mu$. In contrast, $c_\mu$ increases with the contact friction coefficient $\mu_c$.

\subsection{Failure of standard rheology models}
\subsubsection{Failure of viscoplastic rheology and fluidity scaling}
The viscoplastic $\mu(I)$ rheology predicts that there is a general relationship between the friction coefficient $\mu$ and the inertial number $I$ across dense, dry granular flows~\cite{Jopetal06}. However, Figs.~\ref{Scaling2}(a) and \ref{ViscoplasticRheology}(a) show that this prediction fails for our simulated nearly dry flows. This failure cannot be remedied by nonlocal extensions of the $\mu(I)$ rheology. For example, it has been demonstrated that the nonlocal model of Ref.~\cite{KamrinKoval12} is based on a general scaling of the rescaled fluidity $\dot\gamma d/(\mu\sqrt{T})$ with the particle volume fraction $\phi$~\cite{ZhangKamrin17}, which is disobeyed across our dry flows [Fig.~\ref{ViscoplasticRheology}(b)].

\subsubsection{Failure of viscoinertial rheology models}
\begin{figure*}[htb]
 \begin{center}
  \includegraphics[width=2.0\columnwidth]{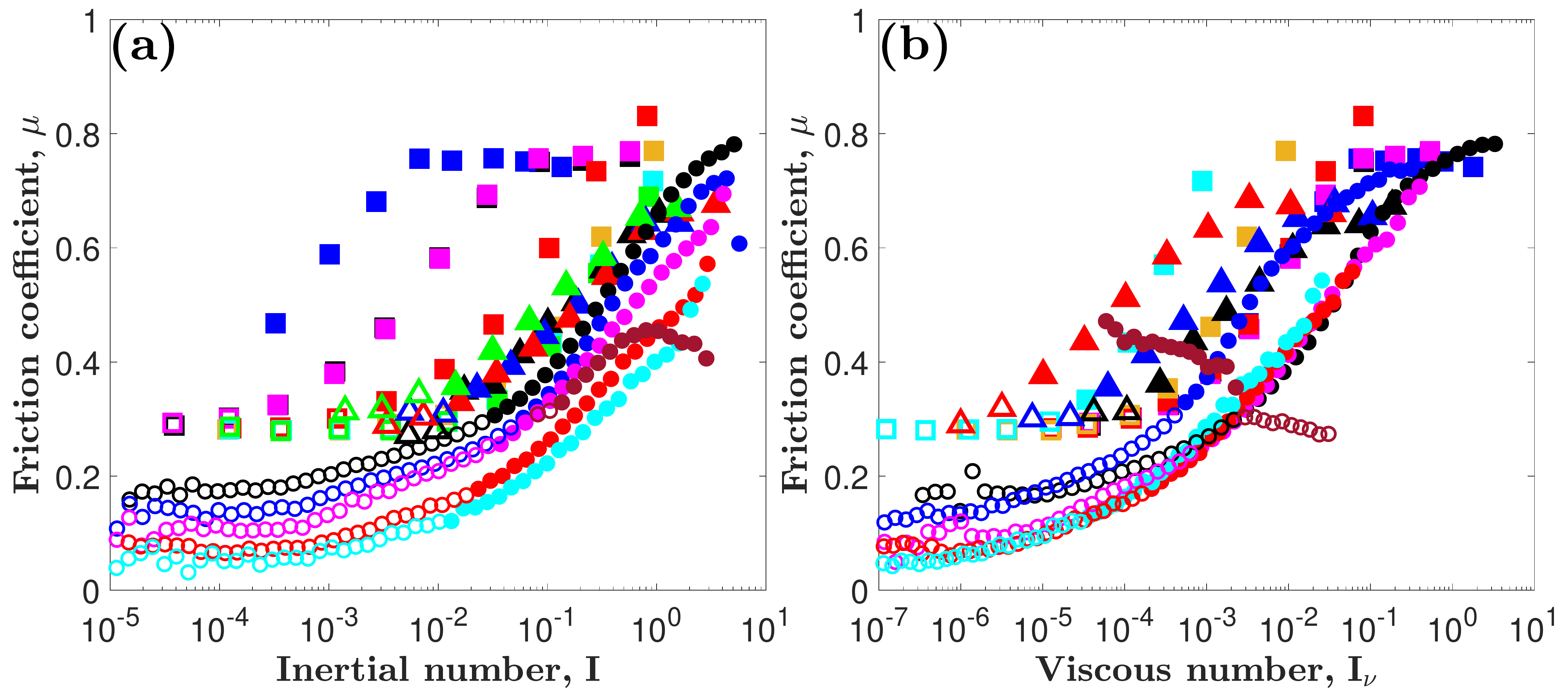}
 \end{center}
 \caption{Mohr-Coulomb friction coefficient $\mu$ vs (a) inertial number $I$ and (b) viscous number $I_\nu$ for data from discrete element method-based simulations of various granular flows. For rotating drum, sediment transport, and gravity flow simulations, $I$, $I_\nu$, and $\mu$ depend on the location within the flow. The value of $\mu$ at each location with $\lambda(\phi)<d$ and either $\mathrm{Pe}<0.49\land M<0.7$ (open symbols) or $\mathrm{Pe}\geq0.49\lor M\geq0.7$ (closed symbols) is allocated to the corresponding bin of $I$ or $I_\nu$. Each bin consists of data from either a single simulation (rotating drum and uniform flows) or from various simulations of the same regime (sediment transport and gravity flows, see Table~\ref{SimulatedConditions}). The mean of $\mu$ within each bin is represented by the symbols (for standard deviation, see Fig.~S2). For the squares, the color order (green, cyan, orange, red, magenta, blue, black) corresponds to $\eta_f/\sqrt{\rho_pP_{zz}d^2}=[0,10^{-3},10^{-2},10^{-1},10^0,10^1,\infty]$. For the triangles, the color order (green, red, blue, black) corresponds to $\eta_f/(\rho_f\omega d^2)=[0,1/160,1/16,3/16]$. For symbol legend, see Fig.~\ref{Scaling1}.}
\label{Scaling2}
\end{figure*}

\begin{figure*}[htb]
 \begin{center}
  \includegraphics[width=2.0\columnwidth]{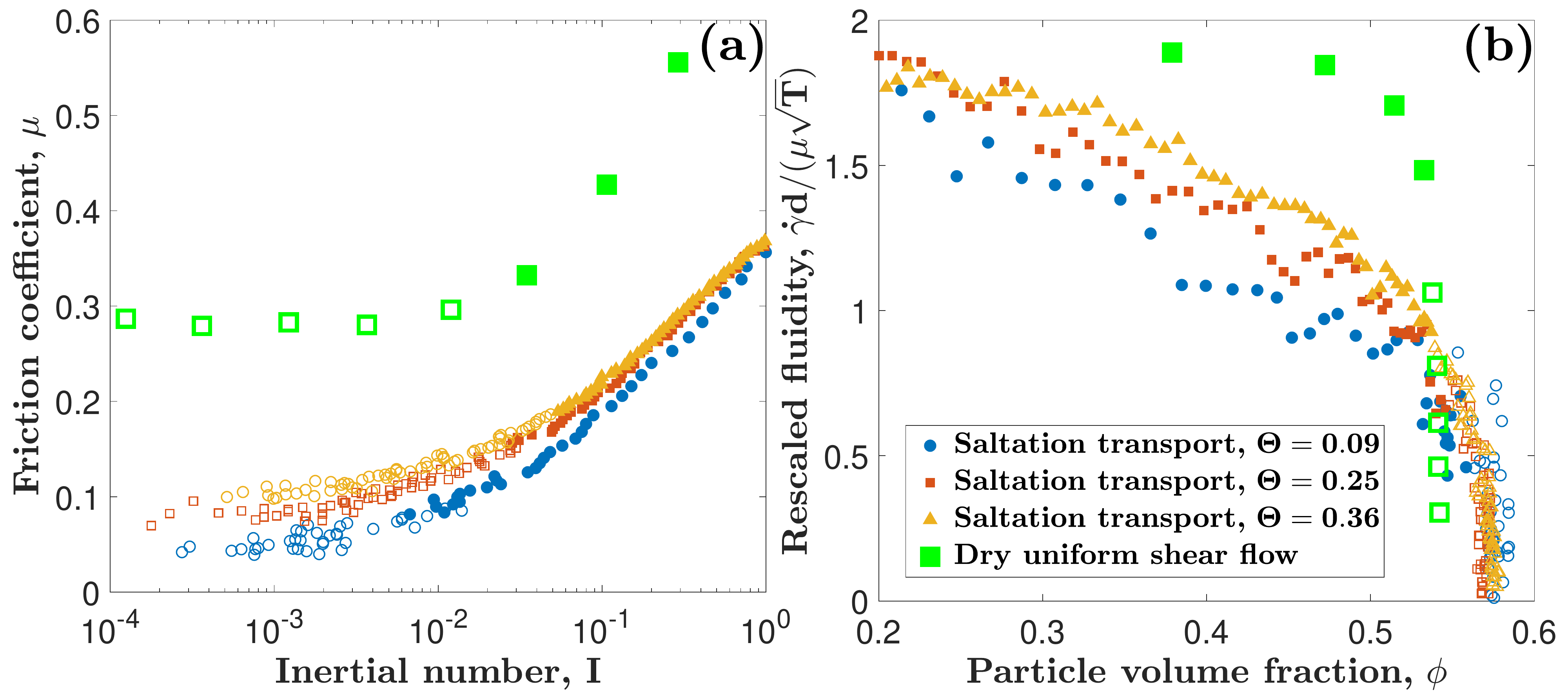}
 \end{center}
 \caption{Failure of viscoplastic rheology and fluidity scaling. (a) Mohr-Coulomb friction coefficient $\mu$ vs inertial number $I$. (b) Rescaled fluidity [$\dot\gamma d/(\mu\sqrt{T})$] vs particle volume fraction $\phi$. The small symbols correspond to data from discrete element method-based simulations of turbulent saltation transport ($s=2000$ and $\mathrm{Ga}=5$) for three different Shields numbers $\Theta$. The large green squares correspond to data from discrete element method-based simulations of dry uniform shear flows. Closed symbols indicate $\mathrm{Pe}\geq0.49\lor M\geq0.7$. Open symbols indicate $\mathrm{Pe}<0.49\land M<0.7$.}
\label{ViscoplasticRheology}
\end{figure*}

\begin{figure*}[htb]
 \begin{center}
  \includegraphics[width=2.0\columnwidth]{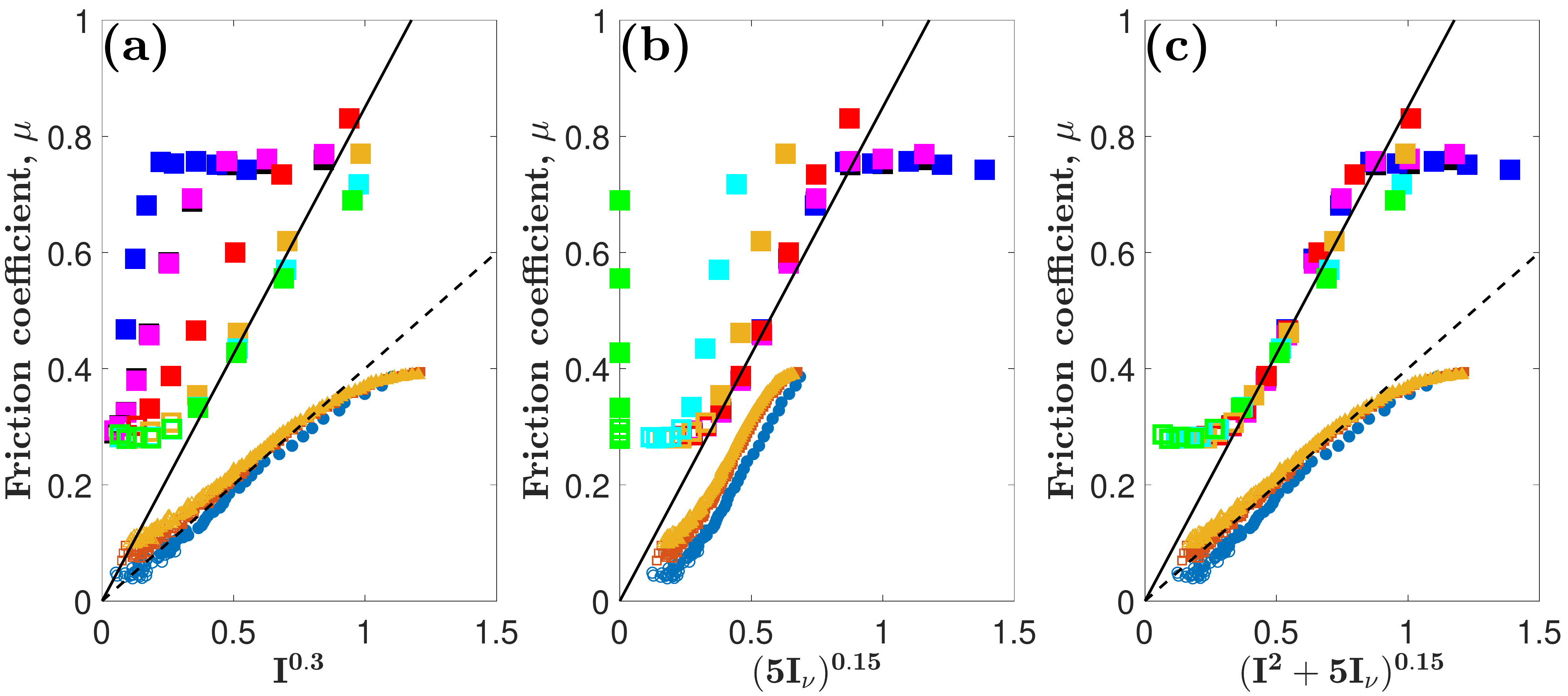}
 \end{center}
 \caption{Failure of viscoinertial rheology. Mohr-Coulomb friction coefficient $\mu$ vs (a) $I^{0.3}$, (b) $(5I_\nu)^{0.15}$, and (c) $(I^2+5I_\nu)^{0.15}$. The small symbols correspond to data from discrete element method-based simulations of turbulent saltation transport ($s=2000$ and $\mathrm{Ga}=5$) for three different Shields numbers $\Theta$. The large squares correspond to data from discrete element method-based simulations of uniformly sheared particle and suspension flows, where the color order (green, cyan, orange, red, magenta, blue, black) indicates $\eta_f/\sqrt{\rho_fP_{zz}d^2}=[0,10^{-3},10^{-2},10^{-1},10^0,10^1,\infty]$. Closed symbols indicate $\mathrm{Pe}\geq0.49\lor M\geq0.7$. Open symbols indicate $\mathrm{Pe}<0.49\land M<0.7$.}
\label{ViscoinertialRheology}
\end{figure*}

Figures~\ref{Scaling2}(b) and \ref{ViscoinertialRheology}(b) show that the failure of the $\mu(I)$ [Figs.~\ref{Scaling2}(a), \ref{ViscoplasticRheology}(a), and \ref{ViscoinertialRheology}(a)] also cannot be remedied by the viscous number rheology $\mu(I_\nu)$~\cite{Boyeretal11}. Interestingly, turbulent saltation transport roughly obeys the same power law with the viscoinertial number $\sqrt{I^2+5I_\nu}$ as uniformly sheared particle and suspension flows [$\mu\propto(I^2+5I_\nu)^{0.15}$], but the proportionality constant is much smaller [Fig.~\ref{ViscoinertialRheology}(c)].

\subsubsection{Failure of extended kinetic theory}
\begin{figure*}[htb]
 \begin{center}
  \includegraphics[width=2.0\columnwidth]{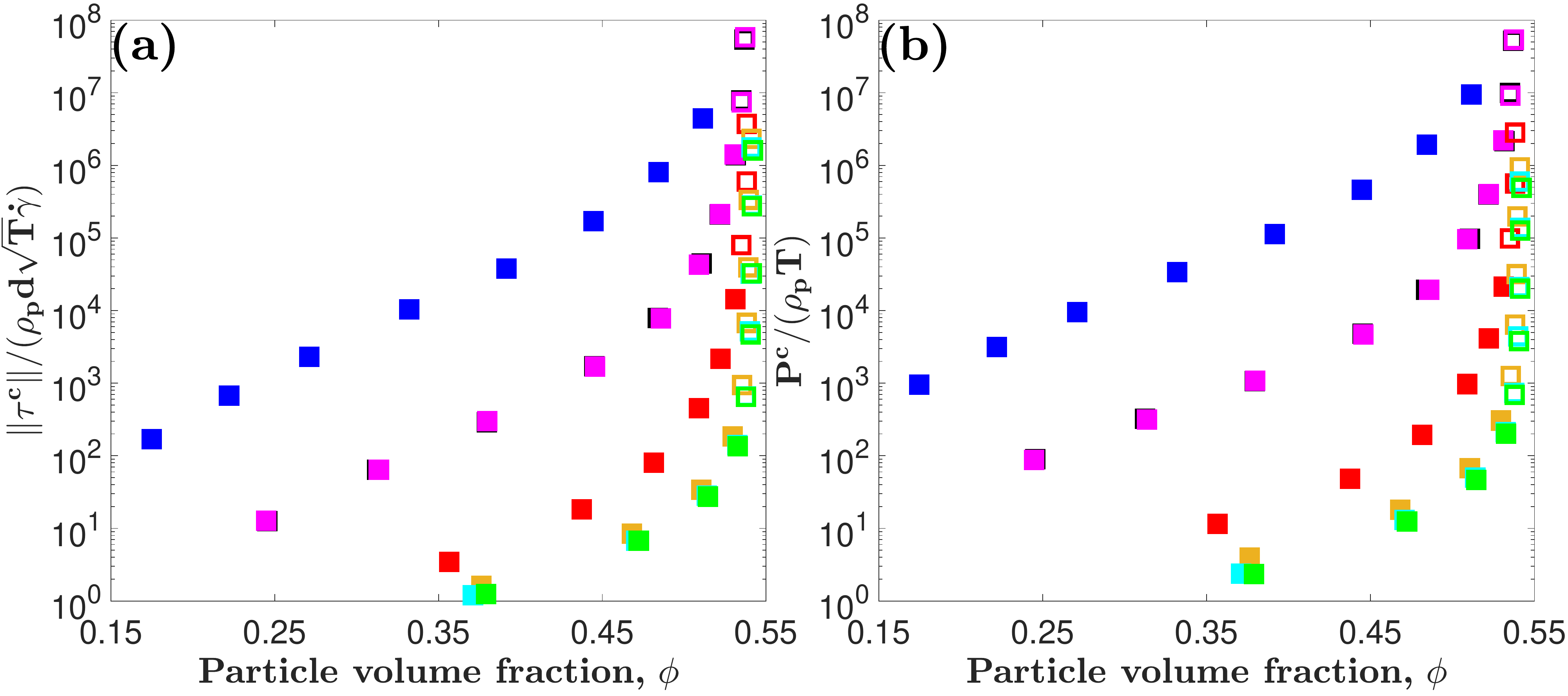}
 \end{center}
 \caption{Failure of extended kinetic theory. (a) Rescaled shear stress [$\lVert\tau^c\rVert/(\rho_pd\sqrt{T}\dot\gamma)$] and (b) rescaled pressure [$P^c/(\rho_pT)$] vs particle volume fraction $\phi$ for data from discrete element method-based simulations of uniformly sheared particle and suspension flows, where the color order (green, cyan, orange, red, magenta, blue, black) indicates $\eta_f/\sqrt{\rho_fP_{zz}d^2}=[0,10^{-3},10^{-2},10^{-1},10^0,10^1,\infty]$. Closed symbols indicate $\mathrm{Pe}\geq0.49\lor M\geq0.7$. Open symbols indicate $\mathrm{Pe}<0.49\land M<0.7$.}
\label{KineticTheory}
\end{figure*}

Extended kinetic theory predicts that the particle shear stress $\lVert\tau^c\rVert\equiv|P^c_1-P^c_2|/2$ (for two-dimensional flows), rescaled by $\rho_pd\sqrt{T}\dot\gamma$, and the rescaled particle pressure $P^c/(\rho_pT)$ are functions of only the particle volume fraction $\phi$. However, it is well known that this prediction, in general, does not hold for wet granular flows~\cite{Garzoetal12,Chamorroetal15,SahaAlam17,Alametal19}, which is shown in Fig.~\ref{KineticTheory} for uniformly sheared particle and suspension flows.

\begin{table*}
    \begin{tabular}{| p{0.28\textwidth} | p{0.15\textwidth} | p{0.19\textwidth} | p{0.17\textwidth} | p{0.16\textwidth} |}
    \hline
     Category & Density ratio $s$ & Galileo number $\mathrm{Ga}$ & Range of $\sqrt{\Theta}$ or $\alpha$ & \# of simulations \\ \hline
		 Viscous bedload transport & $2.65$ & $2$ & $\sqrt{\Theta}\in[0.79,1.39]$ & $11$ \\
		 Viscous bedload transport & $2.65$ & $5$ & $\sqrt{\Theta}\in[0.73,1.18]$ & $10$ \\
		 Viscous bedload transport & $2.65$ & $10$ & $\sqrt{\Theta}\in[0.77,1.12]$ & $8$ \\
		 Viscous bedload transport & $100$ & $0.5$ & $\sqrt{\Theta}\in[0.84,1.02]$ & $4$ \\
		 Viscous bedload transport & $2000$ & $0.1$ & $\sqrt{\Theta}\in[0.84,1.05]$ & $4$ \\
		 \hline
		 Turbulent bedload transport & $2.65$ & $20$ & $\sqrt{\Theta}\in[0.67,0.91]$ & $7$ \\
		 Turbulent bedload transport & $2.65$ & $50$ & $\sqrt{\Theta}\in[0.61,1.45]$ & $19$ \\
		 Turbulent bedload transport & $2.65$ & $100$ & $\sqrt{\Theta}\in[0.6,1.1]$ & $12$ \\
		 \hline
		 Bedload-saltation transition & $100$ & $2$ & $\sqrt{\Theta}\in[0.52,0.7]$ & $8$ \\
		 Bedload-saltation transition & $100$ & $5$ & $\sqrt{\Theta}\in[0.4,0.6]$ & $5$ \\
		 Bedload-saltation transition & $2000$ & $0.5$ & $\sqrt{\Theta}\in[0.73,0.9]$ & $3$ \\
		 \hline
		 Viscous saltation transport & $100$ & $10$ & $\sqrt{\Theta}\in[0.34,0.64]$ & $11$ \\
		 Viscous saltation transport & $2000$ & $2$ & $\sqrt{\Theta}\in[0.34,0.72]$ & $7$ \\
		 \hline
		 Turbulent saltation transport & $100$ & $20$ & $\sqrt{\Theta}\in[0.35,0.65]$ & $4$ \\
		 Turbulent saltation transport & $100$ & $50$ & $\sqrt{\Theta}\in[0.28,0.31]$ & $2$ \\
		 Turbulent saltation transport & $100$ & $100$ & $\sqrt{\Theta}\in[0.28,0.31]$ & $2$ \\
		 Turbulent saltation transport & $2000$ & $5$ & $\sqrt{\Theta}\in[0.3,0.6]$ & $4$ \\
		 Turbulent saltation transport & $2000$ & $10$ & $\sqrt{\Theta}\in[0.35,0.45]$ & $2$ \\
		 \hline
		 Gravity flow in ambient air & $2000$ & $2$ & $\alpha\in[51^\circ,60^\circ]$ & $4$ \\
		 Gravity flow in ambient air & $2000$ & $5$ & $\alpha\in[48^\circ,60^\circ]$ & $5$ \\
		 Gravity flow in ambient air & $2000$ & $10$ & $\alpha\in[42^\circ,60^\circ]$ & $7$ \\
		 Gravity flow in ambient air & $2000$ & $20$ & $\alpha\in[36^\circ,60^\circ]$ & $9$ \\
		 Gravity flow in ambient air & $2000$ & $50$ & $\alpha\in[33^\circ,60^\circ]$ & $10$ \\
		 Gravity flow in ambient air & $2000$ & $100$ & $\alpha\in[30^\circ,60^\circ]$ & $11$ \\
    \hline
    \end{tabular}
		\caption{Summary of the simulated sediment transport and gravity flow conditions included in Figs.~2, S2, and S4.}
		\label{SimulatedConditions}
\end{table*}


\end{document}